\documentclass{article}

\usepackage{microtype}
\usepackage{graphicx}
\usepackage{subfigure}
\usepackage{booktabs} % for professional tables

\usepackage{hyperref}

\usepackage{color,soul}
\usepackage[accepted]{icml2024}

\usepackage{amsmath}
\usepackage{amssymb}
\usepackage{mathtools}
\usepackage{amsthm}
\usepackage{pifont}

\usepackage[capitalize,noabbrev]{cleveref}

\theoremstyle{plain}
\newtheorem{theorem}{Theorem}[section]

\theoremstyle{definition}
\newtheorem{definition}[theorem]{Definition}

\theoremstyle{remark}

\usepackage[textsize=tiny]{todonotes}

\icmltitlerunning{A Space Group Symmetry Informed Network for O(3) Equivariant Crystal Tensor Prediction}

\begin{document}

\twocolumn[
\icmltitle{A Space Group Symmetry Informed Network for O(3) Equivariant Crystal Tensor Prediction}



\icmlsetsymbol{equal_senior}{*}

\begin{icmlauthorlist}
\icmlauthor{Keqiang Yan}{tamu}
\icmlauthor{Alexandra Saxton}{tamu}
\icmlauthor{Xiaofeng Qian}{tamu,equal_senior}
\icmlauthor{Xiaoning Qian}{tamu,equal_senior}
\icmlauthor{Shuiwang Ji}{tamu,equal_senior}
\end{icmlauthorlist}
\icmlaffiliation{tamu}{Texas A\&M University, College Station, TX 77843, USA}
% \icmlaffiliation{tamumat}{Company Name, Location, Country}
% \icmlaffiliation{tamuece}{School of ZZZ, Institute of WWW, Location, Country}
% \icmlaffiliation{equal_senior}{Equal senior authorship}

\icmlcorrespondingauthor{Shuiwang Ji}{sji@tamu.edu}
\icmlcorrespondingauthor{Keqiang Yan}{keqiangyan@tamu.edu}

\icmlkeywords{Machine Learning, ICML}

\vskip 0.3in
]



\printAffiliationsAndNotice{\icmlEqualContribution} % otherwise use the standard text.

\begin{abstract}
We consider the prediction of general tensor properties of crystalline materials,
including dielectric, piezoelectric, and elastic tensors. 
A key challenge here is how to make the predictions satisfy the unique tensor equivariance to $O(3)$ group and invariance to crystal space groups. 
To this end, we propose a General Materials Tensor Network~(GMTNet), which is carefully designed to satisfy the required symmetries. To evaluate our method, we curate a dataset and establish evaluation metrics that are tailored to the intricacies of crystal tensor predictions. Experimental results show that our GMTNet not only achieves promising performance on crystal tensors of various orders but also generates predictions fully consistent with the intrinsic crystal symmetries. Our code is publicly available as part of the AIRS library (\url{https://github.com/divelab/AIRS}).
\end{abstract}
\section{Introduction}

Tensor properties of crystalline materials are fundamental in advancing various technological sectors, leading to significant innovations in device development. These properties span multiple orders, encompassing atomic charge (order 0), atomic force (order 1), dielectric tensor (order 2), piezoelectric tensor (order 3), elastic tensor (order 4), and beyond. Dielectric materials, characterized by dielectric tensors, are crucial in modern technologies ranging from computer memory to sensors and communication circuits~\cite{petousis2016benchmarking}. Piezoelectric materials, notable for their significant piezoelectric coefficients, find extensive use in actuators, sensors, and energy-harvesting devices~\cite{mahapatra2021piezoelectric}. 
Materials with higher-order optical tensors play an essential role in developing novel optics and quantum technologies~\cite{xiao2020berrymemory,wang2020nlpc}.

Accurately predicting these tensor properties is key to discovering new crystals with desirable characteristics. First-principles methods, such as density functional theory (DFT), have facilitated the prediction of various properties within an acceptable error margin compared to traditional laboratory experiments~\cite{petousis2016benchmarking}. However, DFT methods are often resource-intensive, particularly for high-order tensor properties of large crystals, mainly due to the need for self-consistent electronic and ionic relaxation with explicit representations of electronic wavefunctions.

To address these challenges, recent advances have leveraged machine learning (ML) techniques~\cite{graphormer, liu2021spherical, Gong2023Examining}, such as descriptor-based Automatminer~\cite{dunn2020benchmarking} and graph neural networks like MEGNET~\cite{megnet}, for efficient prediction of single-value properties invariant to 3D rotations and translations. Nevertheless, these approaches generally overlook the inherent anisotropy of most crystal systems and the tensorial nature of their macroscopic properties, which are not $E(3)$ invariant and take the form of matrices with dimensions $3^n$, where $n$ denotes the tensor order.

% \shuiwang{description of facts should be in present (not future) tense.}

% \shuiwang{$O(3), E(3)$ in math mode}

% \shuiwang{use $\ell$, instead of $l$}

% Concretely, crystal dielectric tensor $\boldsymbol{\varepsilon} \in \mathbb{R}^{3\times 3}$, via the equation $\mathbf{D} = \boldsymbol{\varepsilon} \mathbf{E}$, determines the electric displacement vector $\mathbf{D} \in \mathbb{R}^3$ (response) when applying an external electrical field $\mathbf{E} \in \mathbb{R}^3$ to a material. When the crystal structure is rotated by $\mathbf{R}$, where $\mathbf{R} \in \mathbb{R}^{3\times3}$, $|\mathbf{R}|=\pm 1$, instead of remaining the same, the dielectric tensor $\boldsymbol{\varepsilon} \in \mathbb{R}^{3\times 3}$ is rotated as $\mathbf{R} \boldsymbol{\varepsilon} \mathbf{R}^T$. Similarly, the piezoelectric tensor $\mathbf{e} \in  \mathbb{R}^{3\times3\times3}$ will rotate to $e_{ijk}' = \sum_{\ell mn} \mathbf{R}_{i\ell}\mathbf{R}_{jm}\mathbf{R}_{kn}e_{\ell mn}$. Thus, ideal ML methods should be able to capture the anisotropy in crystal tensor properties and naturally satisfy this equivariance when predicting crystal tensor properties.

Concretely, the dielectric tensor $\boldsymbol{\varepsilon} \in \mathbb{R}^{3\times 3}$ of a crystal plays a pivotal role in determining its response to external electrical fields. Through the relation $\mathbf{D} = \boldsymbol{\varepsilon} \mathbf{E}$, this tensor dictates the electric displacement vector $\mathbf{D} \in \mathbb{R}^3$ when an electric field $\mathbf{E} \in \mathbb{R}^3$ is applied. Crucially, the dielectric tensor adapts to the crystal's orientation. A rotation of the crystal structure by $\mathbf{R} \in \mathbb{R}^{3\times3}$ with $|\mathbf{R}|=\pm 1$ leads to a corresponding rotation of $\boldsymbol{\varepsilon}$, represented as $\mathbf{R} \boldsymbol{\varepsilon} \mathbf{R}^T$. A similar transformation rule applies to the piezoelectric tensor $\mathbf{e} \in \mathbb{R}^{3\times3\times3}$, where $\mathbf{e}_{ijk}'$ transforms to $\sum_{\ell mn} \mathbf{R}_{i\ell}\mathbf{R}_{jm}\mathbf{R}_{kn}\mathbf{e}_{\ell mn}$. Ideal ML models should inherently capture this anisotropy and these equivariance rules when predicting crystal tensor properties.

% Additionally, the intrinsic symmetries of individual crystals play a critical role in the their tensor properties. Crystals have dramatically different tensor formats in terms of non-zero elements and equality, which are directly dependent on specific crystal class or crystalline point group it belongs to. For example, the dielectric tensor $\boldsymbol{\varepsilon}_{ij}$ of any crystal in the cubic system has non-zero diagonal elements only and three diagonal elements are exactly the same, that is, $\boldsymbol{\varepsilon}_{xx}=\boldsymbol{\varepsilon}_{yy}=\boldsymbol{\varepsilon}_{zz}$ and $\boldsymbol{\varepsilon}_{ij}=0$ for $i \neq j$. In contrast, the dielectric tensor of crystals in the orthorhombic system has non-zero diagonal elements only, but the diagonal elements are independent of each other. While all nine elements of the $3 \times 3$ dielectric tensor are nonzero for crystals in the triclinic system.  

Furthermore, the intrinsic symmetries of crystals significantly influence their tensor properties. Different crystal systems exhibit unique tensor characteristics, determined by their crystal class or space group. For instance, crystals in the cubic system possess dielectric tensors with only non-zero diagonal elements that are identical ($\boldsymbol{\varepsilon}_{xx} = \boldsymbol{\varepsilon}_{yy} = \boldsymbol{\varepsilon}_{zz}$), and zero for off-diagonal elements ($\boldsymbol{\varepsilon}_{ij} = 0$ for $i \neq j$). In contrast, orthorhombic systems feature non-zero diagonal elements that are independent of each other. Triclinic crystals display a more complex dielectric tensor where all nine elements of the $3 \times 3$ matrix are non-zero.

% Given these variations, it is paramount for ML methods to accurately predict tensor properties that reflect the specific symmetries inherent to various crystal structures, ensuring both physical fidelity and relevance to practical applications.

Therefore, for both physical consistency and practical applications, it is essential for ML methods to predict tensor properties that fully respect the intrinsic symmetries of arbitrary crystals. However, how to naturally enforce ML methods to incorporate intrinsic symmetries of crystals and satisfy tensor formats across various crystal systems and space groups is challenging and unsolved.
% Although it is possible to train ML models for individual tensor element without explicitly considering crystal symmetry, such brute-force models can neither guarantee the prediction of all the zero elements, nor guarantee the equality of those mutually dependent elements in the general rank-$n$ tensors consistent with their crystal systems or crystal space groups. 

To address these challenges, our work focuses on the prediction of key crystal tensor properties including dielectric, piezoelectric, and elastic tensors. The contributions of this work are summarized as follows. (1) We propose a general equivariant graph neural network that captures the anisotropy nature and the unique equivariance of crystalline materials tensor properties, GMTNet. (2) GMTNet produces tensor predictions that respect the intrinsic symmetries present in various crystal systems. (3) We have curated a dataset encompassing dielectric, piezoelectric, and elastic tensors, along with establishing robust evaluation metrics specifically tailored for ML-based predictions of general tensor properties. (4) Our work includes detailed proofs and methodological guidance aimed at achieving tensor equivariance and adherence to crystal symmetry constraints, serving as a valuable resource for future research in this field.

\begin{figure*}[t]
    \centering
    \includegraphics[width=0.97\linewidth]{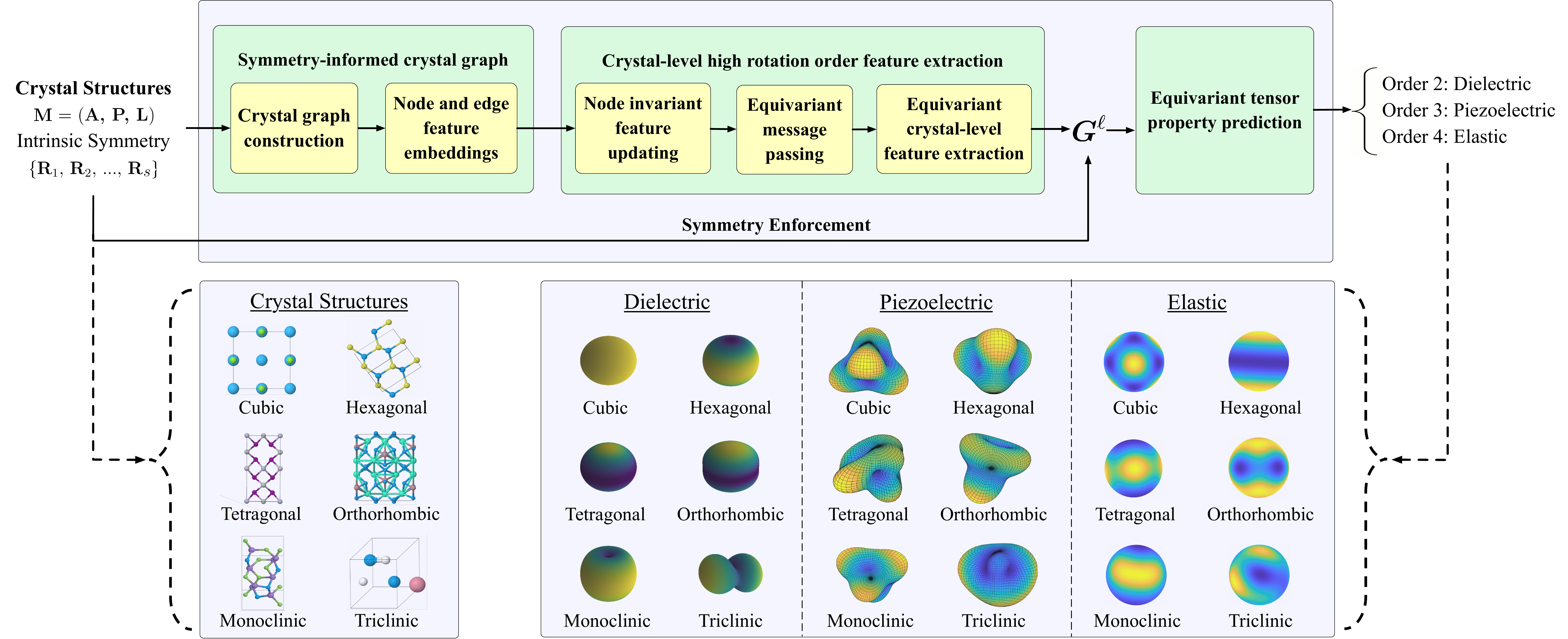}
    \caption{Overview of GMTNet. GMTNet takes crystal structures represented as $\mathbf{M} = (\mathbf{A}, \mathbf{P}, \mathbf{L})$ as input to predict crystal tensor properties of various orders. It comprises four modules: symmetry-informed crystal graph construction, crystal-level equivariant feature extraction, equivariant tensor property prediction, and symmetry enforcement. GMTNet is carefully designed to generate tensor predictions adhere to the intrinsic symmetries of the input crystal structures. We also include visualizations of crystal structures and tensors with different orders belonging to various crystal systems. These visualizations, generated using matplotlib~\cite{matplotlib} and MTEX~\cite{mtex}, illustrate the correlation between crystal symmetries and tensor property complexities.}
    \label{fig:pipeline}
    \vspace{-3mm}
\end{figure*}

\section{Preliminaries and Background}

% In this section, we first briefly describe crystal structures and crystal tensor properties and then demonstrate the influence of $O(3)$ group and crystal space group transformations on crystal tensor properties.

\subsection{Crystal Structures and Tensor Properties}

\textbf{Crystal structures}. Following notations in \citet{yan2022periodic}, a crystal structure is characterized by a unit cell containing a set of atoms, which repeats indefinitely in three-dimensional space along three periodic lattice vectors. It is mathematically represented by $\mathbf{M} = (\mathbf{A}, \mathbf{P}, \mathbf{L})$, where $\mathbf{A}=[\boldsymbol{a}_1, \boldsymbol{a}_2, \cdots, \boldsymbol{a}_n] \in \mathbb{R}^{d_a \times n}$ denotes the $d_a$-dimensional feature vectors of $n$ atoms within the unit cell. $\textbf{P} = [\boldsymbol{p}_1, \boldsymbol{p}_2, \cdots, \boldsymbol{p}_n] \in \mathbb{R}^{3 \times n}$ encapsulates the 3D Euclidean positions of these $n$ atoms. The lattice matrix $\textbf{L} = [\boldsymbol{\ell}_1, \boldsymbol{\ell}_2, \boldsymbol{\ell}_3] \in \mathbb{R}^{3\times 3}$ defines the three periodic lattice vectors, representing the repeating patterns of the unit cell in three-dimensional space. The infinite structure of a given crystal $\mathbf{M} = (\mathbf{A}, \mathbf{P}, \mathbf{L})$ is formalized as
\begin{equation*}
\label{equi_q}
\begin{aligned}
    \hat{\mathbf{P}} = \{\hat{\boldsymbol{p}_i} | \hat{\boldsymbol{p}_i} = \boldsymbol{p}_i + k_1\boldsymbol{\ell}_1 + k_2\boldsymbol{\ell}_2 + k_3\boldsymbol{\ell}_3,~ k_1, k_2, k_3 \in \mathbb{Z}, \\
    i \in \mathbb{Z}, 1 \le i \le n \}, 
    \\
    \hat{\mathbf{A}} = \{\hat{\boldsymbol{a}_i} | \hat{\boldsymbol{a}_i} = \boldsymbol{a}_i, i \in \mathbb{Z}, 1 \le i \le n \},
\end{aligned}
\end{equation*}
where $\hat{\mathbf{P}}$ denotes the positions of atoms and their infinite replicates in the 3D space, and $\hat{\mathbf{A}}$ corresponds to the feature vectors for each atom and its replicas.

\textbf{Crystal tensor properties}. In general, the physical properties are naturally defined as the responses of materials under external fields/perturbations. In this work, we focus on three crystal tensor properties; namely dielectric, piezoelectric, and elastic tensors. The dielectric tensor $\boldsymbol{\varepsilon}$ correlates the externally applied electric field $\mathbf{E} \in \mathbb{R}^{3}$ to the electric displacement field $\mathbf{D} \in \mathbb{R}^{3}$ within the material, expressed as
$\mathbf{D}_i = \sum_j\boldsymbol{\varepsilon}_{ij} \mathbf{E}_j$ with $\boldsymbol{\varepsilon} \in  \mathbb{R}^{3\times3}$ and $i, j \in \{1, 2, 3\}$. The piezoelectric tensor $\mathbf{e} \in  \mathbb{R}^{3\times3\times3}$ relates the applied strain $\boldsymbol{\epsilon} \in \mathbb{R}^{3\times3}$ to the electric displacement field $\mathbf{D}  \in \mathbb{R}^{3}$ within the material, formulated as
$\mathbf{D}_i = \sum_{jk}\mathbf{e}_{ijk} \boldsymbol{\epsilon}_{jk}$ with  $i, j, k \in \{1, 2, 3\}$. Lastly, the elastic tensor $\boldsymbol{C} \in  \mathbb{R}^{3\times3\times3\times3}$ connects the applied strain tensor $\boldsymbol{\epsilon} \in \mathbb{R}^{3\times3}$ to the stress tensor $\boldsymbol{\sigma} \in \mathbb{R}^{3\times3}$ within the material, depicted by
$\boldsymbol{\sigma}_{ij} =  \sum_{k\ell}\boldsymbol{C}_{ijk\ell} \boldsymbol{\epsilon}_{k\ell}$ with $i, j, k, \ell \in \{1, 2, 3\}$. These tensors play pivotal roles in understanding and predicting the mechanical and electrical behavior of crystals under different conditions.

\subsection{Symmetry Constraints of Crystal Tensor Properties}
\label{sym_constraint}

\textbf{$O(3)$ group}. The $O(3)$ group encompasses all rotations and reflections in 3D space. It is mathematically represented by $\mathbf{R} \in \mathbb{R}^{3\times3}$, $|\mathbf{R}|=\pm 1$. 
In the context of crystal tensor properties,  when applying an $O(3)$ group transformation to a crystal structure, from $\mathbf{M} = (\mathbf{A}$, $\mathbf{P}$, $\mathbf{L})$ to $\mathbf{M}' = (\mathbf{A}$, $\mathbf{R}\mathbf{P}$, $\mathbf{R}\mathbf{L})$, the properties of the crystal tensor change accordingly. This change can be mathematically described as $\boldsymbol{\varepsilon}' = \mathbf{R} \boldsymbol{\varepsilon} \mathbf{R}^T$. In practical terms, this means that the relationship between an externally applied electric field $\mathbf{E} \in \mathbb{R}^{3}$ and the resultant electric displacement field $\mathbf{D} \in \mathbb{R}^{3}$, which is typically defined as $\mathbf{D} = \boldsymbol{\varepsilon} \mathbf{E}$, will be altered under the rotation transformation. Specifically, the electric field and electric displacement field transform to $\mathbf{R}\mathbf{E}$ and $\mathbf{R}\mathbf{D}$, respectively. This leads to a new relationship as $\mathbf{R}\mathbf{D} = \boldsymbol{\varepsilon}' \mathbf{R}\mathbf{E}$. Rearranging this equation, we obtain $\mathbf{D} = \mathbf{R}^T \boldsymbol{\varepsilon}' \mathbf{R}\mathbf{E}$, which confirms that the transformed dielectric tensor after rotation is $\boldsymbol{\varepsilon}' = \mathbf{R} \boldsymbol{\varepsilon} \mathbf{R}^T$. This transformation principle extends similarly to other crystal tensors. For instance, the piezoelectric tensor transforms as $\mathbf{e}_{ijk}' = \sum_{\ell mn} \mathbf{R}_{i\ell}\mathbf{R}_{jm}\mathbf{R}_{kn}\mathbf{e}_{\ell mn}$, and the elastic tensor follows the transformation $\boldsymbol{C}_{ijk\ell}' = \sum_{pqrs} \mathbf{R}_{ip}\mathbf{R}_{jq}\mathbf{R}_{kr}\mathbf{R}_{\ell s}\boldsymbol{C}_{pqrs}$.

\textbf{Crystal space group}. Crystal space group transformations encapsulate the spatial symmetries inherent in crystal structures. Mathematically, these transformations can be represented by a rotation-reflection matrix $\mathbf{R} \in \mathbb{R}^{3\times3}$, where $|\mathbf{R}|=\pm 1$, combined with a translation vector $\mathbf{b} \in \mathbb{R}^{3}$, implying that the transformation of $\mathbf{M} = (\mathbf{A}$, $\mathbf{P}$, $\mathbf{L})$ to $\mathbf{M}' = (\mathbf{A}$, $\mathbf{R}\mathbf{P} + \mathbf{b}$, $\mathbf{L})$ transforms the inner unit cell structure back to itself, albeit with a reindexed arrangement. 
% It is worth noting that $\mathbf{R}\mathbf{L} = \mathbf{L}$ for $\mathbf{R}$ in crystal symmetry group, and $\mathbf{R}\mathbf{P} + \mathbf{b}$ transforms the inner unit cell structure back to itself as shown in Figure.~\ref{fig:1}.

By Neumann's Principle, a fundamental tenet in crystal physics, the symmetry observed in any physical property of a crystal
% , such as the dielectric tensor, 
must mirror the spatial symmetry of the crystal structure itself. This principle suggests that the response of a crystal property to an external perturbation, such as an electric field, maintains its symmetry under space group transformations. For instance, considering the dielectric tensor, when an arbitrary electric field $\mathbf{E} \in \mathbb{R}^{3}$ is subjected to any transformation $\mathbf{R}$ within the crystal's space group, the resulting electric displacement field $\mathbf{D}$ transforms accordingly, maintaining the symmetry as $\mathbf{R}\mathbf{D} = \boldsymbol{\varepsilon} \mathbf{R}\mathbf{E}$. This indicates that rotating the external electric field by an operation of the space group yields a correspondingly rotated response in the electric displacement field.
By using $\mathbf{R}_i\mathbf{D} = \boldsymbol{\varepsilon} \mathbf{R}_i\mathbf{E}$ where $\mathbf{R}_i$ is in the crystal's space group, it can be seen  $\mathbf{D} = \mathbf{R}_i^T \boldsymbol{\varepsilon} \mathbf{R}_i\mathbf{E}$, and $\boldsymbol{\varepsilon} = \mathbf{R}_i^T \boldsymbol{\varepsilon} \mathbf{R}_i$, or equivalently $\boldsymbol{\varepsilon} = \mathbf{R}_i \boldsymbol{\varepsilon} \mathbf{R}_i^T$. This relationship imposes specific constraints on the elements of the $\boldsymbol{\varepsilon}$ matrix, leading to the presence of zero elements and multually dependent elements in certain positions of the matrix, with a detailed demonstration provided in Appendix~\ref{app:demonstration}.

In summary, crystal tensor properties change accordingly when rotating or reflecting the crystal structure and have symmetry constraints, e.g., $\boldsymbol{\varepsilon} = \mathbf{R}_i \boldsymbol{\varepsilon} \mathbf{R}_i^T$ with $\mathbf{R}_i$ in the crystal's space group for dielectric tensors. These constraints are the direct consequence of symmetry, which is not only true for \emph{all} crystal tensor properties, but also critical to be obeyed when developing ML-based tensor property predictions.

\section{Method}

In this section, we describe our material tensor network, termed GMTNet, aiming at predicting crystal tensor properties. GMTNet is designed to address two fundamental questions: (1) How can a ML pipeline be structured to ensure that the predicted tensors adapt appropriately under $O(3)$ transformations across various tensor properties? (2) How can symmetry constraints be integrated into the pipeline to ensure that the output tensors inherently adhere to the corresponding constraints? To initiate this discussion, we first establish a formal definition of crystal tensor property prediction task, followed by the demonstration of GMTNet.

\begin{definition}[Crystal Tensor Property Prediction] 
\label{def:1}
The task of crystal tensor property prediction involves predicting the Tensor properties of crystals, such as the dielectric tensor $\boldsymbol{\varepsilon} \in \mathbb{R}^{3\times 3}$, piezoelectric tensor $\mathbf{e} \in  \mathbb{R}^{3\times3\times3}$, and elastic tensor $\boldsymbol{C} \in  \mathbb{R}^{3\times3\times3\times3}$. These properties are predicted in their tensor matrix forms, using the crystal structure $\mathbf{M} = (\mathbf{A}$, $\mathbf{P}$, $\mathbf{L})$ as the input.
\end{definition} 

GMTNet presents a general end-to-end machine learning solution tailored for the prediction of crystal tensor properties. It consists of four core modules: a symmetry-informed crystal graph construction module, a crystal-level equivariant feature extraction module, an equivariant tensor property prediction module, and a symmetry enforcement module. The overall framework is visually represented in Fig.~\ref{fig:pipeline}.

\subsection{Symmetry-informed Crystal Graph Construction}
\label{sec:31}

To predict crystal tensor properties adhere to intrinsic symmetries, we introduce the symmetry-informed crystal graph construction module. This module is responsible for transforming a given crystal structure $\mathbf{M} = (\mathbf{A}$, $\mathbf{P}$, $\mathbf{L})$ into a corresponding crystal graph representation.

\textbf{Requirements for atom-level features}. As elucidated in Sec.~\ref{sym_constraint}, crystal symmetry underscores the interconnected relationships between atoms within the crystal. Specifically, for a space group transformation in the crystal defined by $\mathbf{R} \in \mathbb{R}^{3\times3}$, $|\mathbf{R}|=\pm 1$, and a translation vector $\mathbf{b} \in \mathbb{R}^{3}$, if atom $i$ in $\mathbf{M}$ is mapped to atom $j$ in the transformed structure $\mathbf{M}' = (\mathbf{A}$, $\mathbf{R}\mathbf{P} + \mathbf{b}$, $\mathbf{L})$, then atoms $i$ and $j$ must be of the same type (an $E(3)$ invariant feature) and exhibit correspondingly rotated forces (an $O(3)$ equivariant feature). Therefore, atom-level features extracted via machine learning should also adhere to these physical symmetry constraints to ensure accurate and meaningful predictions.

\textbf{Crystal graph construction}. To satisfy the above requirements for atom-level features, we construct crystal graphs in the following manner. For a given crystal structure $\mathbf{M} = (\mathbf{A}, \mathbf{P}, \mathbf{L})$, where $\mathbf{A}=[\boldsymbol{a}_1, \boldsymbol{a}_2, \cdots, \boldsymbol{a}_n] \in \mathbb{R}^{d_a \times n}$, we create $n$ nodes representing atoms and their periodic duplicates in 3D space. Each node $i$ is associated with node feature $\boldsymbol{a}_i$ and positions $\hat{\boldsymbol{p}_i}$, defined as $\boldsymbol{p}_i + k_1\boldsymbol{\ell}_1 + k_2\boldsymbol{\ell}_2 + k_3\boldsymbol{\ell}_3$, with $k_1, k_2, k_3 \in \mathbb{Z}$.  The neighbors of each node are determined within a radius $r$, established by the distance to the $k$-th nearest neighbor. Edges are formed between nodes within this radius, with edge features capturing the relative positions. Concretely, given the $k$-th nearest neighbor $m$ with position $\boldsymbol{p}_m$, $r = \text{Euclidean}(\boldsymbol{p}_m - \boldsymbol{p}_i)$, and for any node $j$ with position $\boldsymbol{p}_j'=\boldsymbol{p}_j + k_1\boldsymbol{\ell}_1 + k_2\boldsymbol{\ell}_2 + k_3\boldsymbol{\ell}_3$ that satisfies $\text{Euclidean}(\boldsymbol{p}_j' - \boldsymbol{p}_i) \le r$, an edge will be built from node $j$ to $i$ with edge feature $\boldsymbol{\nu}_{ji}=\boldsymbol{p}_j' - \boldsymbol{p}_i$. If there are multiple positions of node $j$ within the radius, multiple edges will be built from $j$ to $i$.  This graph construction method ensures compliance with atom-level feature requirements.

\textbf{Node and edge feature embedding}. Following previous works~\cite{alignn, yan2022periodic}, node type features are embedded into 92-dimensional CGCNN~\cite{cgcnn} feature representations. Edge features $\boldsymbol{\nu}_{ji}$ are transformed into a combination of their magnitude, $||\boldsymbol{\nu}_{ji}||_2$, and normalized direction, $\hat{\boldsymbol{\nu}}_{ji}$. The magnitude is further mapped to a potential-like term, $-c/||\boldsymbol{\nu}_{ji}||_2$, encoded using radial basis function (RBF) kernels, as suggested in previous works~\cite{potnet}.

\subsection{Crystal-level Equivariant Feature Extraction}
\label{sec:global_extract}
\label{sec:32}

Focusing on crystal tensor properties such as dielectric, piezoelectric, and elastic tensors, we introduce the crystal-level equivariant feature extraction module. This module is responsible for extracting crystal-level equivariant features from crystal graphs. These features are pivotal for the subsequent prediction of tensor properties.

\textbf{Requirements for crystal-level features}. In line with Neumann's Principle, it is essential that the extracted crystal-level features retain the same spatial symmetry characteristics as the original crystal structure. This alignment ensures that the features accurately reflect the inherent symmetries of the crystal, which is crucial for the precise prediction of tensor properties.

\textbf{Node invariant feature updating}. Node invariant features are updated using the state-of-the-art Comformer~\cite{comformer} invariant layers. Specifically, messages are transmitted from a neighboring node $j$ to node $i$ by utilizing node features ($\boldsymbol{f}_j$, $\boldsymbol{f}_i$) and edge feature ($\boldsymbol{f}_{ji}^e$); subsequently, these messages from all neighbors are aggregated to update $\boldsymbol{f}_i$. The message from node $j$ to $i$ is formed by the query $\boldsymbol{q}_{ji}= \text{LN}_Q(\boldsymbol{f}_i)$, key $\boldsymbol{k}_{ji}=(\text{LN}_K(\boldsymbol{f}_i) | \text{LN}_K(\boldsymbol{f}_j))$, and value features $\boldsymbol{v}_{ji}=(\text{LN}_V(\boldsymbol{f}_i) | \text{LN}_V(\boldsymbol{f}_j) | \text{LN}_E(\boldsymbol{f}_{ji}^e))$, where $\text{LN}_Q, \text{LN}_K, \text{LN}_V, \text{LN}_E$ denote the linear transformations. We derive the corresponding message by the following operations: 
\begin{equation}
\begin{aligned}
    \boldsymbol{\alpha}_{ji} = & \frac{ \boldsymbol{q}_{ji} \star  \boldsymbol{\xi}_{K}(\boldsymbol{k}_{ji})}{\sqrt{d_{\boldsymbol{q}_{ji}}}}, \\
    \boldsymbol{msg}_{ji} =  & \text{sigmoid}(\text{BN}(\boldsymbol{\alpha}_{ji})) \star 
    \boldsymbol{\xi}_{V}(\boldsymbol{v}_{ji}),
\end{aligned}
\end{equation}
where $\boldsymbol{\xi}_{K}, \boldsymbol{\xi}_{V}$ represent nonlinear transformations applied to key and value features, and the operators $\star$ and $|$ denote the Hadamard product and concatenation. $\text{BN}$ refers to the batch normalization layer, and $d_{\boldsymbol{q}_{ji}}$ indicates the dimensionality of $\boldsymbol{q}_{ji}$. 
Then, node feature $\boldsymbol{f}_i$ is updated as follows,
\begin{equation}
    \boldsymbol{msg}_{i} = \sum_{j \in \mathcal{N}_i} \boldsymbol{msg}_{ji}, ~ \boldsymbol{f}_i^{\text{new}} = \boldsymbol{\xi}_{msg}(\boldsymbol{f}_i + \text{BN}(\boldsymbol{msg}_{i})),
\end{equation}
with $\boldsymbol{\xi}_{msg}$ denoting the softplus activation function.

\textbf{Equivariant message passing}. We then obtain atom-level high rotation order features ($\ell > 0$) by the widely-used tensor field network (TFN)~\cite{thomas2018tensor}, renowned for its efficacy in achieving 3D rotation, reflection, and permutation equivariance. Following conventions of \citet{QHNet}, the TFN layer employs the tensor product $\otimes$ to amalgamate two irreducible representations, $u$ and $v$, each characterized by distinct rotation orders $\ell_1$ and $\ell_2$. This fusion process leverages the Clebsch-Gordan (CG) coefficients~\cite{CGcoeffi}, resulting in a new irreducible representation with rotational order $\ell_3$:
\begin{equation}
    (u^{\ell_1} \otimes v^{\ell_2})_{m_3}^{\ell_3} = \sum_{m_1=-\ell_1}^{\ell_1} \sum_{m_2=-\ell_1}^{\ell_1} \text{CG}_{(\ell_1, m_1), (\ell_2, m_2)}^{(\ell_3, m_3)} u_{m_1}^{\ell_1} v_{m_2}^{\ell_2},
\end{equation}
where the CG matrix is denoted as CG, $\ell \in \mathbb{N}$, with the condition $|\ell_1 - \ell_2| \le \ell_3 \le \ell_1 + \ell_2$, and $m \in \mathbb{N}$ represents the $m$-th element in the irreducible representation, constrained by $-\ell \le m \le \ell$. Analogous to the TFN, our proposed equivariant layer comprises both filter and convolution modules. Within the filter module, spherical harmonic filters $Y$ are applied to the directional edge feature  $\hat{\boldsymbol{\nu}}_{ji}$ and integrated with the invariant edge feature $\boldsymbol{f}_{ji}^e$ to form messages from node $j$ to node $i$ represented as $F$ as,
\begin{equation}
    F_{c,m}^{(\ell_{\text{in}}, \ell_f)} (\boldsymbol{f}_{ji}^e, \hat{\boldsymbol{\nu}}_{ji}) = \boldsymbol{\xi}_{c}^{(\ell_\text{in}, \ell_f)} (\boldsymbol{f}_{ji}^e) Y_m^{\ell_f}(\hat{\boldsymbol{\nu}}_{ji}),
\end{equation}
where $\boldsymbol{\xi}_{c}$ represents a nonlinear layer corresponding to channel index $c$. The convolutional module then aggregates neighboring messages to the center node $i$ through tensor product as,
\begin{equation}
    \boldsymbol{f}_{i,c}^{\ell_\text{out}}=\sum_{j \in \mathcal{N}_i}(F^{(\ell_{\text{in}, c}, \ell_f)} (\boldsymbol{f}_{ji}^e, \hat{\boldsymbol{\nu}}_{ji}) \otimes \boldsymbol{f}_i^{\ell_{\text{in}}})^{\ell_\text{out}},
\end{equation}
adhering to the constraint  $|\ell_{\text{in}} - \ell_f| \le \ell_\text{out} \le \ell_{\text{in}} + \ell_f$.
The stacking of multiple layers of this mechanism allows for the extraction of high rotation order features at the atomic level.

\textbf{Equivariant crystal-level feature extraction}. After obtaining high rotation order equivariant node features, the crystal-level equivariant features are aggregated as follows,
\begin{equation}
    \boldsymbol{G}^{\ell} = \frac{1}{n} \sum_{1 \le i \le n} \boldsymbol{f}_{i}^{\ell}.
\end{equation}

A detailed analysis confirming that these crystal-level equivariant features meet the pre-established requirements is presented in Appendix~\ref{app:proof}.

\subsection{Equivariant Tensor Property Prediction}
\label{sec:33}

Constructing tensor predictions such as dielectric, piezoelectric, and elastic tensors in matrix forms that satisfy all crystal symmetry constraints and $O(3)$ equivariance is non-trivial. Rather than directly employing crystal-level equivariant features to construct matrices, our approach simulates the physical responses that these properties represent. 

Let's take dielectric tensor $\boldsymbol{\varepsilon}$ as an example. It characterizes the electric displacement field response $\mathbf{D} \in \mathbb{R}^3$ of a crystal under an external electric field $\mathbf{E} \in \mathbb{R}^3$, with $\mathbf{D} = \boldsymbol{\varepsilon} \mathbf{E}$. 
We predict an $O(3)$ equivariant electric displacement $\mathbf{D}$ by treating $\mathbf{E}$ as a conditional input and utilizing tensor products. This approach effectively simulates the interaction between the electrical field $\mathbf{E}$ and the crystal. The implementation employs a tensor product layer as 
\begin{equation}
    \mathbf{D} = \sum_{\ell_G}(\boldsymbol{G}^{\ell_G} \otimes \mathbf{E}^{\ell_{\mathbf{E}}=1})^{\ell_{\mathbf{D}}=1},
\end{equation}
where $\mathbf{E}$ and $\mathbf{D}$ are vectors of rotation order $\ell=1$. The dielectric tensor is subsequently derived by calculating the gradient $\boldsymbol{\varepsilon} = \frac{\partial \mathbf{D}}{\partial \mathbf{E}}$.

This approach ensures that the predicted dielectric tensor inherently adheres to the required equivariance properties. The versatility of this module extends to other crystal tensor properties, each conforming to their unique equivariance criteria. The corresponding modules for piezoelectric and elastic tensors are developed as follows:
\begin{itemize}
    \item Piezoelectric tensor: strain $\boldsymbol{\epsilon}$ is a $3\times3$ tensor that consists of irreducible representations with rotation order 0, 1, and 2. The corresponding tensor product is $\mathbf{D} = \sum_{\ell_G} \sum_{\ell_\epsilon} (\boldsymbol{G}^{\ell_G} \otimes \boldsymbol{\epsilon}^{\ell_\epsilon})^{\ell_{\mathbf{D}}=1}$, and $\mathbf{e} = \frac{\partial \mathbf{D}}{\partial \boldsymbol{\epsilon}}$.
    \item Elastic tensor: strain $\boldsymbol{\epsilon}$ and stress $\boldsymbol{\sigma}$ are $3\times3$ tensor that consist of irreducible representations with rotation order 0, 1, and 2. The corresponding tensor product is $\boldsymbol{\sigma}^{\ell_\sigma} = \sum_{\ell_G} \sum_{\ell_\epsilon} (\boldsymbol{G}^{\ell_G} \otimes \boldsymbol{\epsilon}^{\ell_\epsilon})^{\ell_\sigma}$, and $C = \frac{\partial \boldsymbol{\sigma}}{\partial \boldsymbol{\epsilon}}$.
\end{itemize}

\subsection{Crystal Symmetry Enforcement Module}

It is worth noting that GMTNet \textbf{already satisfies both $O(3)$ tensor equivariance and space group constraints} using components demonstrated in Sec.~\ref{sec:31}, Sec.~\ref{sec:32}, and Sec.~\ref{sec:33}, including symmetry-informed crystal graph construction, crystal-level equivariant feature extraction, and equivariant tensor property prediction. Additionally, we aim to build a robust system that satisfies $O(3)$ tensor equivariance and space group constraints not only for ideal crystal inputs and message passing without numerical errors but also for realistic crystal inputs and message passing operations with small errors.

\textbf{Challenges in crystal symmetry}. In the materials science field, the determination of crystal symmetry in crystalline structures is often subject to a distortion tolerance. For large-scale crystal datasets, such as the Materials Project~(MP) and JARVIS, a typical Euclidean distance tolerance is set at $0.01$. This means that if the largest pairwise distortion, measured before and after a symmetry transformation, falls below this threshold, the transformation is considered part of the crystal's symmetry group. Hence, minor distortions in the input crystal structures can disrupt the crystal symmetry. Additionally, numerical errors during message passing will also slightly alter the pairwise relationships between atoms, as outlined in the atom-level feature requirements in Sec.~\ref{sec:global_extract}. These distortions and numerical inaccuracies present challenges in generating predictions that fully comply with crystal symmetry constraints.

\textbf{Crystal symmetry enforcement module}. To address the challenge of upholding crystal symmetry in tensor predictions, we introduce a crystal symmetry enforcement module. This module simplifies the complex symmetry constraints of tensor properties into constraints applicable to crystal-level features. For any transformation $\mathbf{R} \in \mathbb{R}^{3\times3}, \mathbf{b} \in \mathbb{R}^{3}$ within the crystal's symmetry group, applying $\mathbf{M}' = (\mathbf{A}$, $\mathbf{R}\mathbf{P} + \mathbf{b}$, $\mathbf{L})$ leaves the crystal structure unchanged. However, the node features for each node $i$ will be modified as follows:
\begin{equation}
     {\boldsymbol{f}_{i}^{\ell}}' = \text{WD}^{\ell} (\mathbf{R}) \circ \boldsymbol{f}_{i}^{\ell},
\end{equation}

where $\circ$ denotes matrix multiplication and $\text{WD}^{\ell}(\mathbf{R})$ denotes the Wigner D-transformation matrix for the irreducible representation with rotation order $\ell$ and 3D rotation $\mathbf{R}$. The crystal-level representation then becomes 
\begin{equation}
     {\boldsymbol{G}^{\ell}}' = \frac{1}{n} \sum_{1 \le i \le n} \text{WD}^{\ell} (\mathbf{R}) \circ \boldsymbol{f}_{i}^{\ell} = \text{WD}^{\ell} (\mathbf{R})  \circ \boldsymbol{G}^{\ell}.
\end{equation}
For a crystal structure belonging to a specific crystal space group with transformations $\{\mathbf{R}_1, \mathbf{R}_2, \cdots, \mathbf{R}_s\}$, due to the fact that macroscopic tensor properties of crystals will not change under translations, we focus on the rotation and reflection transformations and remove duplicates to form $\{\mathbf{R}_1, \mathbf{R}_2, \cdots, \mathbf{R}_{n_{r}}\}$. We verify that this set still constitutes a group and adheres to the four group conditions. To impose symmetry constraints on tensor predictions, we ensure that:
$\boldsymbol{G}^{\ell} = \text{WD}^{\ell} (\mathbf{R})  \circ \boldsymbol{G}^{\ell},$
for every $\mathbf{R}$ in $\{\mathbf{R}_1, \mathbf{R}_2, \cdots, \mathbf{R}_{n_{r}}\}$ by setting
\begin{equation}
     \boldsymbol{G}^{\ell}_{sym} = \frac{1}{{n_{r}}}\sum_{1\le i \le {n_{r}}}\text{WD}^{\ell} (\mathbf{R}_i)  \circ \boldsymbol{G}^{\ell},
\end{equation}
and group theory guarantees that:
\begin{equation}
     \text{WD}^{\ell} (\mathbf{R}_m) \circ \boldsymbol{G}^{\ell}_{sym} = \boldsymbol{G}^{\ell}_{sym}.
\end{equation}
Additionally, we can reduce small distortions in the input crystal structures by refining the crystal input structures using their space group transformations.

\subsection{Equivariance and Symmetry Verification}
\label{sec:analysis}

\textbf{Equivariance of tensor properties}. Applying $\mathbf{R} \in \mathbb{R}^{3\times 3}, |\mathbf{R}| = \pm 1$ to the crystal structure and the external electrical field will result in ${\boldsymbol{G}^{\ell}}'=\text{WD}^{\ell} (\mathbf{R}) \boldsymbol{G}^{\ell}$, and 
\begin{equation*}
    \mathbf{R}\mathbf{D} = \sum_{\ell_G}(\text{WD}^{\ell_G} (\mathbf{R}) \circ \boldsymbol{G}^{\ell_G} \otimes \mathbf{R}\mathbf{E}^{\ell_{\mathbf{E}}=1})^{\ell_{\mathbf{D}}=1}.
\end{equation*}
The corresponding $\boldsymbol{\varepsilon}'$ will be
\begin{equation*}
    \boldsymbol{\varepsilon}' = \frac{\partial (\mathbf{R}\mathbf{D})}{\partial (\mathbf{R}\mathbf{E})} = \mathbf{R} \frac{\partial \mathbf{D}}{\partial \mathbf{E}}\mathbf{R}^{-1} = \mathbf{R} \boldsymbol{\varepsilon} \mathbf{R}^{T},
\end{equation*}
which satisfies the tensor's equivariance, and similar properties can be proven for piezoelectric and elastic tensors.

\textbf{Crystal symmetry of tensor properties}. Let's consider a pair of  external perturbation and crystal response ($\mathbf{E}, \mathbf{D}$). When the  crystal structure is rotated by $\mathbf{R}_i$ (one of the crystal's space group symmetry operations), while $\mathbf{E}$ remains constant, the resultant displacement field $\mathbf{D}'$ is given by
\begin{align*}
    \mathbf{D}' = & \sum_{\ell_G}(\text{WD}^{\ell_G} (\mathbf{R}_i) \circ \boldsymbol{G}^{\ell_G} \otimes \mathbf{E}^{\ell_{\mathbf{E}}=1})^{\ell_{\mathbf{D}}=1} \\
    = & \sum_{\ell_G}(\boldsymbol{G}^{\ell_G} \otimes \mathbf{E}^{\ell_{\mathbf{E}}=1})^{\ell_{\mathbf{D}}=1} = \mathbf{D},
\end{align*}
due to $\text{WD}^{\ell_G} (\mathbf{R}_i) \circ \boldsymbol{G}^{\ell_G} = \boldsymbol{G}^{\ell_G}$, with the dielectric tensor $\boldsymbol{\varepsilon}'$ being
\begin{equation*}
    \boldsymbol{\varepsilon}' = \frac{\partial \mathbf{D}'}{\partial \mathbf{E}} = \frac{\partial \mathbf{D}}{\partial \mathbf{E}} = \boldsymbol{\varepsilon},
\end{equation*}
after the transformation $\mathbf{R}_i$. Similar proofs apply to the piezo and elastic tensors, underscoring the consistency of these tensor properties with crystal symmetry principles.

\section{Related Works}

\textbf{Equivariant graph neural networks}. Equivariant graph neural networks have demonstrated remarkable capabilities in modeling atomic systems~\cite{schnet, zhang2023artificial} including molecules~\cite{egnn, schutt2021equivariant, liao2022equiformer, nequip, mace}, materials~\cite{cdvae, luo2023towards,comformer}, and proteins~\cite{gvp_protein,fu2023latent}. These networks have been  particularly developed to predict either invariant scalar properties such as formation energy and band gap, or equivariant 3D vector properties including atomic forces and positional shifts. Despite these advancements, the specific challenge of predicting high-order tensor properties such as dielectric, piezoelectric, and elastic tensors while maintaining their inherent equivariance properties remains unexplored. Furthermore, the task of ensuring that tensor predictions conform to the stringent crystal symmetry constraints presents a unique and complex challenge that has not been addressed in previous works.

\textbf{Machine learning based crystal property prediction}. The realm of crystal property prediction has witnessed a substantial leap forward with the advent of machine learning techniques. These methods~\cite{cgcnn,megnet,gatgnn,alignn,yan2022periodic,deng2023chgnet,coGN} offer a significant acceleration that are often several orders of magnitude faster than traditional Density Functional Theory (DFT) calculations while maintaining a commendable level of prediction accuracy. However, the primary focus of most existing studies has been on predicting scalar properties. Extending beyond scalar property prediction, \citet{m3gnet} introduced a method enabling invariant networks to predict atomic forces and stress by deriving gradients relative to energy predictions. Yet, it falls short to predict a broader range of tensor properties inherent in crystals.

Recent work by \citet{etgnn} has furthered this progress, enabling invariant graph neural networks to produce tensor-form predictions through the outer product of edge vectors. This technique generates a $3\times 3$ matrix from $\boldsymbol{\nu}_{ji} \boldsymbol{\nu}_{ji}^T$ and a $3 \times 3 \times 3$ matrix from $\boldsymbol{\nu}_{ji} \cdot \boldsymbol{\nu}_{ji} \cdot \boldsymbol{\nu}_{ji}$, where $\cdot$ denotes the outer product. Despite its innovation, this approach introduces a significant computational complexity of $O(nk^{m-1})$ for tensors of order-$m$ with $k$ denoting the average number of edges per atom, and requires specialized designs for different tensors to account for anti-symmetry (e.g., $\boldsymbol{\nu}_{ji} \boldsymbol{\nu}_{ji}^T$ does not include anti-symmetry). Most notably, it fails to enforce crystal symmetry constraints in tensor predictions as shown in the experimental results.

In contrast, our method significantly diverges from \citet{etgnn}.  We achieve a consistent computational complexity of $O(nk)$, with only two-hop message passing for tensors of varying orders, and inherently resolves the issue of anti-symmetry often encountered in outer product operations like $\boldsymbol{\nu}_{ji} \boldsymbol{\nu}_{ji}^T$.  Critically, our method integrates crystal symmetry constraints that are the direct consequence of symmetry and critical to be obeyed when developing ML-based tensor property predictions. This has been demonstrated in our experimental results. 
Our innovative technique represents a notable stride forward in the field of machine learning-based prediction of crystal tensor properties, effectively overcoming the limitations of previous methods and enhancing the scope and accuracy of predictive modeling in the domain of materials science.

Additionally, instead of carefully designing model components to achieve the invariance of crystal space groups when predicting general tensors with different orders, there is a widely used general approach, group averaging~\cite{GA_murphy2018janossy,GA_yarotsky2022universal}, or more generally, frame averaging~\cite{puny2021frame}, that can convert any model output to be invariant to a group of transformations of interest. However, we show in Appendix~\ref{app:proof} that directly converting outputs to satisfy crystal space groups will result in sub-optimal prediction performances.

\section{Experiments}

\subsection{Curated Crystal Tensor Property Dataset}

In our research, a dataset is curated specifically focusing on crystal tensor properties, including dielectric, piezoelectric, and elastic tensors, sourced from the JARVIS-DFT database~\cite{jarvis}. This dataset has been constructed with a keen emphasis on ensuring congruence between the properties and structures, achieved by extracting both the tensor property values and corresponding crystal structures directly from the DFT calculation files. This approach guarantees that the symmetry of the properties aligns with that of the structures. Notably, each tensor property within this dataset is computed using a consistent DFT core, ensuring uniformity in the calculation method.

\begin{table}[h!]
    \vspace{-3mm}
  \caption{Dataset statistics. Fnorm denotes Frobenius norm.  \label{dataset_sta}}\vspace{1mm}
  \centering
\scalebox{0.65}{
  \begin{tabular}{l|c|c|c|c|c}
    \toprule
     Dataset & \# Samples & Fnorm Mean & Fnorm STD & \# Elem. & Unit\\
    \midrule
    Dielectric&  $4713$& $14.7$ & $18.2$ & $87$ & Unitless\\
    Piezo&  $4998$& $0.43$& $3.09$ & $87$ & C/m$^2$\\
    Elastic &  $14220$& $327$ & $249$ & $87$ & GPa\\
    \bottomrule
  \end{tabular}
  \vspace{-3mm}
}
\end{table}

The dataset's statistics, as detailed in Table~\ref{dataset_sta}, underscores the significant challenges posed in predicting crystal tensor properties. These challenges stem from several factors: (1) the diversity of constituting elements in each dataset, with more than 80 different elements included, (2) the limited number of available training samples, which is less than 5,000 for dielectric and piezoelectric tensors and less than 15,000 for elastic tensors, and (3) the substantial variability observed in the properties, as indicated by the Frobenius norm (denoted as Fnorm in the table). The dielectric tensor is relative dielectric constant with respect to vacuum permittivity $\varepsilon_0$ and unitless ($\varepsilon_0$ = 8.854$\times$10$^{-12}$ CV$^{-1}$m$^{-1}$). 

\begin{figure*}[t!]
    \centering
    \includegraphics[width=0.90\linewidth]{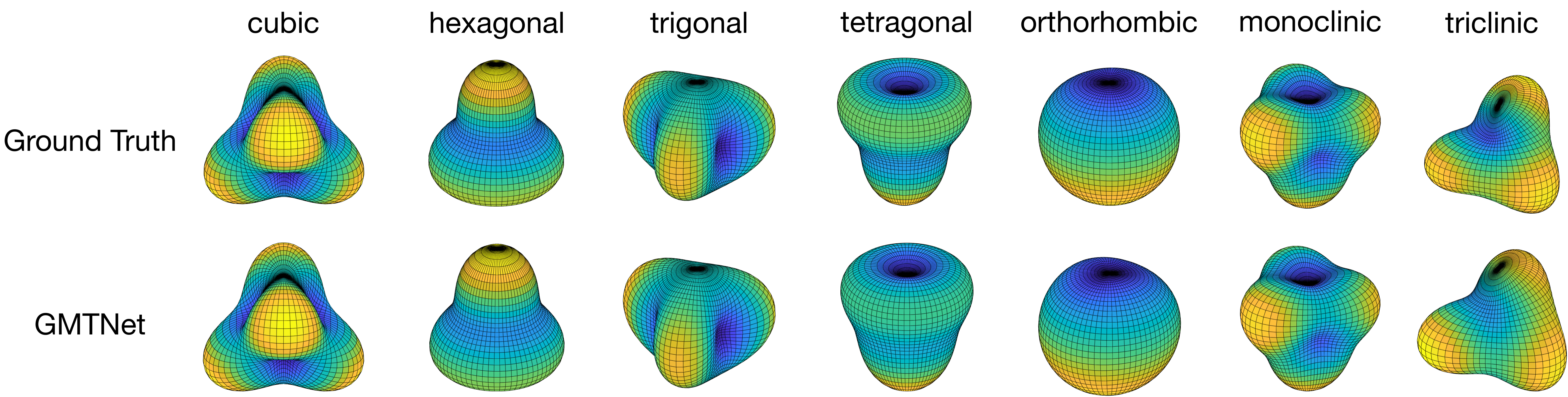}
    \caption{Visualization of piezoelectric tensor predictions on piezoelectric test set. The results  underscore GMTNet's effectiveness in generating symmetry-consistent piezoelectric tensor predictions with tensor order 3.}
    \label{fig:vis_compare_piezo}
    \vspace{-3mm}
\end{figure*}

\subsection{Experimental Setup}
\label{sec:settings}

\textbf{Baseline methods}. We benchmark against two key methods in the field. The first is the invariant MEGNET model~\cite{megnet}, which has been previously employed for predicting dielectric tensors~\cite{morita2020modeling}. The second is ETGNN~\cite{etgnn}, known for its capacity to generate tensor-form predictions.

\textbf{Evaluation metrics for tensor predictions}. Given the limited number of studies capable of producing crystal tensor properties of various orders, there is a need for well-defined evaluation metrics to foster advancements in ML-based tensor property prediction. To thoroughly assess the quality of tensor predictions, we propose the following metrics, each targeting a specific aspect of the tensor prediction quality. (1) \textbf{Success rate in capturing zero elements} that evaluates the ability of the model to correctly identify tensor elements that should be zero-valued due to symmetry constraints inherent in crystal structures. (2) \textbf{Success rate in identifying mutually dependent elements} that measures the model's accuracy in capturing the equality between tensor elements that are dependent due to symmetry.
(3) \textbf{Frobenius norm (Fnorm) distance} that is an $E(3)$ invariant measure for crystal tensor properties, averaged across different crystal systems. It is worth noting that the widely used MAE distance is not $E(3)$ invariant for tensor properties. (4) \textbf{High-quality prediction rate (EwT)} that is determined by the ratio of Fnorm(error) to Fnorm(label) being less than 25\%. This threshold is aligned with DFPT~\cite{petousis2016benchmarking} standards for comparing against experimental results, offering a robust benchmark for prediction accuracy.

\textbf{Experimental settings}. A single NVIDIA A100 GPU is used for computing. We directly follow \citet{megnet} and \citet{etgnn} to implement MEGNET and ETGNN, with more details shown in Appendix~\ref{app:exp}. For each property, we split the samples into training, evaluation, and test sets using ratio 8:1:1. To train our model, we use Huber loss~\cite{huber1992robust} with AdamW~\cite{loshchilov2018decoupled}, $10^{-5}$ weight decay, and polynomial decay for the learning rate. It is worth noting that the Wigner D matrix calculation provided in e3nn~\cite{e3nngeiger2022} introduces numerical errors, and the error is larger for higher rotation order features. To address this issue, we provide a detailed approach with the tolerance that can further adjust the tensor predictions in Appendix~\ref{app:exp}. We use maximum rotation order $\ell_{max}=3$ features for dielectric and piezoelectric tensors, while use $\ell_{max}=4$ for elastic tensors.

\subsection{Experimental Results}

\textbf{Ability to generate tensors adhering to symmetry}. We first evaluate the ability of GMTNet to generate tensor predictions that adhere to crystal symmetry. As demonstrated in Sec.~\ref{sym_constraint}, crystal symmetry constraints result in exact zero elements in dielectric tensors in all crystal systems except the triclinic system. Table~\ref{tab:diel_zero} shows that invariant MEGNET has no ability to capture this, while equivariant ETGNN can only capture a limited portion of crystal structures in various crystal systems. In contrast, our method achieves \textbf{100\%} success rate in capturing zero elements adhering to crystal symmetry, across various crystal systems. 

\begin{table}[h!]
    \vspace{-3mm}
  \caption{Predicting symmetry-constrained zero-valued dielectric tensor elements. Success rate measured by error $<10^{-5}$. Property labels in the curated dataset achieve 100\% success rate.}
  \tiny
  \label{tab:diel_zero}
  \centering
  \begin{tabular}{l|cc|c}
    \toprule
     Crystal System & MEGNET & ETGNN & GMTNet  \\
    \midrule
    Cubic&  $0\%$& $13.5\%$ & $\mathbf{100\%}$ \\
    Tetragonal&  $0\%$& $1.3\%$& $\mathbf{100\%}$ \\
    Hexa-Trigonal &  $0\%$& $2.3\%$ & $\mathbf{100\%}$ \\ 
    Orthorhombic&  $0\%$& $0\%$ & $\mathbf{100\%}$ \\
    Monoclinic & $0\%$& $6.4\%$ & $\mathbf{100\%}$ \\
    \bottomrule
  \end{tabular}
\end{table}

% \begin{table}[h!]
%   \caption{Capturing equality for symmetry informed positions. Success rate measured by difference $<1e^{-4}$.}
%   \tiny
%   \label{main_com_verify}
%   \centering
%   \begin{tabular}{l|c|c|c|c|c}
%     \toprule
%      Crystal System & dataset & MEGNet & w/o constraint & w/o equality & Ours \\
%     \midrule
%     Cubic& $100\%$ & $0\%$& $11\%$ & $100\%$ & $100\%$ \\
%     Tetragonal& $100\%$ & $0\%$& $47.4\%$ & $0\%$ & $100\%$ \\
%     Hexa- and Trigonal & $100\%$ & $0\%$& $24.2\%$ & $0\%$ & $100\%$ \\
%     Monoclinic & $100\%$ & $100\%$& $0\%$ & $100\%$ & $100\%$ \\
%     \bottomrule
%   \end{tabular}
% \end{table}

\begin{table}[h!]
  \caption{Capturing the equality of symmetry-constrained mutually dependent dielectric tensor elements. Success rate measured by difference $<10^{-4}$. Dataset labels achieve 100\% success rate.}
  \tiny
  \label{tab:diel_equal}
  \centering
  \begin{tabular}{l|cc|c}
    \toprule
     Crystal System &  MEGNET & ETGNN & GMTNet \\
    \midrule
    Cubic&  $0\%$& $100\%$ & $\mathbf{100\%}$ \\
    Tetragonal&  $0\%$& $55.3\%$ & $\mathbf{100\%}$ \\
    Hexagonal &  $0\%$& $0\%$& $\mathbf{100\%}$ \\
    Trigonal & $0\%$& $0\%$& $\mathbf{100\%}$ \\
    \bottomrule
  \end{tabular}
  \vspace{-3mm}
\end{table}

Additionally, intrinsic crystal symmetries result in mutually dependent elements in dielectric tensors in certain crystal systems. Table~\ref{tab:diel_equal} demonstrates that MEGNET has \textbf{0\%} success rate for all systems. ETGNN achieves 100\% for cubic systems, but only achieves 55\% for the tetragonal system, and even \textbf{0\%} for hexagonal and trigonal systems, while our method again achieves \textbf{100\%} success rate across various crystal systems. We also provide \textbf{equivariance verification} in Appendix~\ref{app:exp} showing that ETGNN and our GMTNet maintain equivariance for tensor predictions while MEGNET breaks this property.

\begin{table}[h!]
    \vspace{-3mm}
  \caption{Comparison of accuracy in terms of Fnorm and error within threshold (EwT) on the test set of dielectric tensors.}
  \tiny
  \label{frobe_diele}
  \centering
  \begin{tabular}{l|cc|c}
    \toprule
       & MEGNET & ETGNN & GMTNet \\
    \midrule
    Fnorm $\downarrow$& $4.16$& $3.92$ & $\mathbf{3.50}$ \\
    EwT 25\% $\uparrow$ & $74.9\%$& $81.3\%$ & $\mathbf{84.5\%}$ \\
    EwT 10\% $\uparrow$ & $38.9\%$ & $41.6\%$ & $\mathbf{57.1\%}$ \\
    EwT 5\% $\uparrow$ & $19.1\%$ & $23.8\%$ & $\mathbf{27.8\%}$ \\
    % EwT 2\%& - & - & $9.8\%$ & $9.8\%$ \\
    \bottomrule
  \end{tabular}
  \vspace{-2mm}
\end{table}

\textbf{High-quality tensor predictions}. The prediction accuracy of different models by Fnorm and EwT is shown in Table~\ref{frobe_diele}. GMTNet achieves $\mathbf{84.5\%}$ high-quality dielectric predictions within $25\%$ threshold and achieves $57.1\%$ for a stricter $10\%$ threshold, showing a much better modeling power beyond ETGNN with only $41.6\%$. Additionally, GMTNet achieves a 3.50 Fnorm across different crystal systems, significantly better than ETGNN with 3.92. We also provide \textbf{visualization comparison} for predicted piezoelectric tensors in Figure~\ref{fig:vis_compare_piezo}, as well as dielectric and elastic tensors in Appendix~\ref{app:exp}.

\begin{table}[h!]
    \vspace{-4mm}
  \caption{Prediction accuracy in terms of Fnorm and error within threshold (EwT) on test set for piezoelectric and elastic tensors.}
  \small
  \label{frobe_piezo_elastic}
  \centering
  \begin{tabular}{l|c|c}
    \toprule
      & Piezo (C/$m^2$) & Elastic (GPa) \\
    \midrule
    Data Fnorm (mean $\pm$ std) & $0.43 \pm 3.09$& $326.9 \pm 249.3$  \\
    \midrule
    Fnorm $\downarrow$& $\mathbf{0.37}$& $\mathbf{67.38}$  \\
    EwT 25\% $\uparrow$ & $49.1\%$& $66.1\%$ \\
    EwT 10\% $\uparrow$ & $46.3\%$ & $21.8\%$ \\
    EwT 5\% $\uparrow$ & $45.7\%$ & $7.7\%$ \\
    \midrule
    % Symmetry-Zero $\uparrow$ & $100(96.4)\%$ & $100(99.7)\%$ \\
    Symmetry-Zero $\uparrow$ & $100\%$ & $100\%$ \\
    Symmetry-Equality $\uparrow$ & $100\%$ & $100\%$ \\
    \bottomrule
  \end{tabular}
\end{table}

\textbf{Generality of GMTNet for higher order tensors}. We extend the evaluation of GMTNet to \textbf{higher-order tensor} properties, specifically focusing on piezoelectric (order 3) and elastic (order 4) tensors, to showcase the versatility of GMTNet as a comprehensive framework capable of handling a diverse range of tensor properties. The results, as detailed in Table~\ref{frobe_piezo_elastic}, are indicative of GMTNet's robust performance. It consistently achieves a 100\% success rate in accurately identifying zero elements and mutually dependent elements. Furthermore, GMTNet demonstrates impressive proficiency for these more complex tensors, achieving high-quality prediction rates of $49.1\%$ for piezoelectric tensors and $66.1\%$ for elastic tensors, underscoring its effectiveness and general applicability in the domain of tensor property prediction within materials science.

% $14.7 \pm 18.2$

\begin{table}[h!]
    \vspace{-4mm}
  \caption{Efficiency analysis on the dielectric dataset.}
  \tiny
  \label{tab_effi}
  \centering
  \begin{tabular}{l|c|c|c}
    \toprule
    Model  & Time/Batch (s)$\downarrow$ & Num. Param.$\downarrow$ & Fnorm $\downarrow$\\
    \midrule
    ETGNN & 0.121 & 1.1 M & 3.92  \\
    GMTNet & \textbf{0.069 (57\%)} & \textbf{0.7 M (64\%)} & \textbf{3.50}  \\
    \bottomrule
  \end{tabular}
  \vspace{-2mm}
\end{table}

\textbf{Efficiency comparison}. The results presented in Table~\ref{tab_effi} highlight the efficiency gains achieved by our method when compared to ETGNN. Our approach not only reduces the running time by 43\% but also utilizes 37\% fewer trainable parameters,  remarkably accompanied by a significant enhancement in accuracy. These outcomes demonstrate the robustness and superior modeling capability of GMTNet, establishing it as a more efficient and effective solution in the realm of ML-based crystal tensor prediction.

\begin{table}[h!]
    \vspace{-3mm}
  \caption{Ablation studies on the dielectric dataset.}
  \tiny
  \label{tab_ablation}
  \centering
  \begin{tabular}{c|c|c|c|c}
    \toprule
     Equivariance & Struct Correction & Symm. Constraints & Equality & Fnorm \\
    \midrule
    \ding{55} & \ding{55} & \ding{55} & \ding{55} & 4.15 \\
    \checkmark & &  & & 3.76   \\
    \checkmark & \checkmark &  & & 3.72 \\
    \checkmark & \checkmark & \checkmark &  & 3.56 \\
    \checkmark & \checkmark & \checkmark & \checkmark & \textbf{3.50} \\
    \bottomrule
  \end{tabular}
\end{table}

\textbf{Ablation study}. Here, we evaluate the importance of each component in GMTNet. 
To begin with, the equivariant tensor prediction module enables the $O(3)$ equivariant prediction of tensor properties. As shown in Table~\ref{tab_ablation}, 
it plays a vital role and decreases Fnorm from 4.15 to 3.76, which is already better than ETGNN. Without this module, GMTNet cannot provide equivariant predictions.

Furthermore, the symmetry enforcement module, including structure correction and symmetry constraint crystal-level feature correction enhances GMTNet's robustness beyond ideal crystal inputs and message-passing operations for realistic crystal inputs and message passing with numerical errors. It can be seen in Table~\ref{tab_ablation} that the symmetry enforcement module further decreases Fnorm from 3.76 to 3.56, and tolerance-guided prediction adjustment (equality) described in Sec.~\ref{sec:settings} decreases Fnorm to 3.50.

We also demonstrate that GMTNet without the symmetry enforcement module already satisfies $O(3)$ equivariance and space group invariance when predicting crystal tensors for ideal crystal structures, as shown in Table~\ref{tab:verify_1} and Table~\ref{tab:verify_2}. Specifically, we include the zero and equal entry tests on ideal crystal inputs without distortions as shown below. \textbf{GMTNet w/o correction} indicates GMTNet without the symmetry enforcement module. Note that error ratio for zero entries are calculated as $\sum_{ij \in \text{zero positions}} abs(\varepsilon_{ij}) / \sum_{mn \in \text{nonzero positions}} abs(\varepsilon_{mn})$, to measure the relative errors. It can be seen that for ideal crystal inputs, GMTNet w/o correction is more robust in satisfying space group invariance. 

\begin{table}[h!]
    \vspace{-3mm}
  \caption{Predicting symmetry-constrained zero-valued dielectric tensor elements for ideal crystal structures.}
  \tiny
  \label{tab:verify_1}
  \centering
  \begin{tabular}{l|c|c}
    \toprule
     Error ratio $\downarrow$ & ETGNN  &  GMTNet w/o correction   \\
    \midrule
    Cubic& $5.2*10^{-8}$  & \textbf{0}  \\
    Tetragonal&   $5.6*10^{-8}$  & $\mathbf{1.3*10^{-16}}$   \\
    Hexa-Trigonal & $4.3*10^{-3}$ & $\mathbf{5.3*10^{-9}}$  \\ 
    Orthorhombic & $2.5*10^{-8}$ & \textbf{0}  \\
    Monoclinic & $3.5*10^{-8}$ & \textbf{0} \\
    \bottomrule
  \end{tabular}
  \vspace{-3mm}
\end{table}

\begin{table}[h!]
    \vspace{-3mm}
  \caption{Predicting symmetry-constrained equal-valued dielectric tensor elements for ideal crystal structures.}
  \small
  \label{tab:verify_2}
  \centering
  \begin{tabular}{l|c|c}
    \toprule
     Crystal System & ETGNN  &  GMTNet w/o correction  \\
    \midrule
    Cubic& \checkmark & \checkmark \\
    Tetragonal&  \checkmark & \checkmark  \\
    Hexa-Trigonal & \ding{55} & \checkmark  \\
    \bottomrule
  \end{tabular}
\end{table}

Additionally, it is worth noting that GMTNet without the symmetry enforcement module is not robust enough for realistic crystal inputs with structural errors, or in other words, minor distortions that disrupt the crystal symmetry, as shown in Table.~\ref{tab:verify_3}.

\begin{table}[h!]
    \vspace{-3mm}
  \caption{Predicting symmetry-constrained zero-valued dielectric tensor elements. Success rate measured by error $<10^{-5}$. Property labels in the curated dataset achieve 100\% success rate.}
  \tiny
  \label{tab:verify_3}
  \centering
  \begin{tabular}{l|c|cc}
    \toprule
     Crystal System & ETGNN & GMTNet w/o correction & GMTNet  \\
    \midrule
    Cubic& $13.5\%$ & $44.6\%$ &$\mathbf{100\%}$ \\
    Tetragonal& $1.3\%$& $56.6\%$ &$\mathbf{100\%}$ \\
    Hexa-Trigonal & $2.3\%$ & $57.0\%$ &$\mathbf{100\%}$ \\ 
    Orthorhombic& $0\%$ & $68.7\%$ &$\mathbf{100\%}$ \\
    Monoclinic & $6.4\%$ & $78.2\%$ &$\mathbf{100\%}$ \\
    \bottomrule
  \end{tabular}
  \vspace{-3mm}
\end{table}

\section{Conclusion and Limitations}

To conclude, we present GMTNet, a symmetry informed equivariant network for crystal tensor property prediction. It is carefully designed to satisfy the unique equivariance to $O(3)$ group and invariance to crystal space groups, across various tensor properties. A crystal tensor dataset is curated with specifically designed evaluation metrics to foster the advancements in ML-based tensor property prediction. Experimental results demonstrate that GMTNet achieves promising performance on crystal tensors of various orders, and generates predictions fully consistent with the intrinsic crystal symmetries. The limitations of our current GMTNet include (1) the scale of our curated tensor dataset can be expanded, which could potentially enhance the model's robustness and the diversity of its applications, and (2) GMTNet currently cannot generate tensor predictions for amorphous materials. These directions can be explored as future works.

\newpage
\section*{Acknowledgments}
We thank Tian Xie, Chenru Duan, Yuanqi Du, Haiyang Yu, and Youzhi Luo for insightful discussions. X.F.Q. acknowledges partial support from National Science Foundation under awards CMMI-2226908 and DMR-2103842. X.N.Q. acknowledges partial support from National Science Foundation under awards DMR-2119103 and IIS-2212419.
S.J. acknowledges partial support from National Science Foundation under awards IIS-2243850 and CNS-2328395.
\section*{Impact Statement}

This work can be used to discover novel materials with desirable tensor properties. Hence, if misused, the societal consequences of discovering materials with specific tensor properties may apply to this work.

\bibliography{example_paper}
\bibliographystyle{icml2024}

%%%%%%%%%%%%%%%%%%%%%%%%%%%%%%%%%%%%%%%%%%%%%%%%%%%%%%%%%%%%%%%%%%%%%%%%%%%%%%%
%%%%%%%%%%%%%%%%%%%%%%%%%%%%%%%%%%%%%%%%%%%%%%%%%%%%%%%%%%%%%%%%%%%%%%%%%%%%%%%
% APPENDIX
%%%%%%%%%%%%%%%%%%%%%%%%%%%%%%%%%%%%%%%%%%%%%%%%%%%%%%%%%%%%%%%%%%%%%%%%%%%%%%%
%%%%%%%%%%%%%%%%%%%%%%%%%%%%%%%%%%%%%%%%%%%%%%%%%%%%%%%%%%%%%%%%%%%%%%%%%%%%%%%
\newpage
\appendix
\onecolumn
\section{Appendix}

\subsection{Demonstration of the influence of crystal space group on tensor properties}
\label{app:demonstration}

\textbf{Demonstration}. Formally, $\mathbf{D} = \boldsymbol{\varepsilon} \mathbf{E}$ can be written as
\begin{equation*}
    \begin{bmatrix}
    D_x \\
    D_y \\
    D_z
    \end{bmatrix} = 
    \begin{bmatrix}
    \boldsymbol{\varepsilon}_{xx} & \boldsymbol{\varepsilon}_{xy} & \boldsymbol{\varepsilon}_{xz} \\
    \boldsymbol{\varepsilon}_{yx} & \boldsymbol{\varepsilon}_{yy} & \boldsymbol{\varepsilon}_{yz} \\
    \boldsymbol{\varepsilon}_{zx} & \boldsymbol{\varepsilon}_{zy} & \boldsymbol{\varepsilon}_{zz} \\
    \end{bmatrix}
    \begin{bmatrix}
    E_{x} \\
    E_{y} \\
    E_{z} \\
    \end{bmatrix},
\end{equation*}
where each $\boldsymbol{\varepsilon}_{ji}$ element quantifies the impact of the electric field along axis $i$ on the electric displacement along axis $j$. Consider, for instance, a 90-degree rotation around the $z$-axis, denoted as $\mathbf{R}_1$ as
\begin{equation*}
    \mathbf{R}_1 =
    \begin{bmatrix}
    0 & -1 & 0 \\
    1 & 0 & 0 \\
    0 & 0 & 1 \\
    \end{bmatrix}.
\end{equation*}
Applying this rotation to the transformation equation $\boldsymbol{\varepsilon} = \mathbf{R}_i \boldsymbol{\varepsilon} \mathbf{R}_i^T$, we obtain
\begin{equation*} 
    \boldsymbol{\varepsilon}
    = \begin{bmatrix}
    0 & -1 & 0 \\
    1 & 0 & 0 \\
    0 & 0 & 1 \\
    \end{bmatrix} 
    \begin{bmatrix}
    \boldsymbol{\varepsilon}_{xx} & \boldsymbol{\varepsilon}_{xy} & \boldsymbol{\varepsilon}_{xz} \\
    \boldsymbol{\varepsilon}_{yx} & \boldsymbol{\varepsilon}_{yy} & \boldsymbol{\varepsilon}_{yz} \\
    \boldsymbol{\varepsilon}_{zx} & \boldsymbol{\varepsilon}_{zy} & \boldsymbol{\varepsilon}_{zz} \\
    \end{bmatrix}
    \begin{bmatrix}
    0 & 1 & 0 \\
    -1 & 0 & 0 \\
    0 & 0 & 1 \\
    \end{bmatrix}.
\end{equation*}
 Simplifying the above equation yields:
\begin{equation*} 
    \begin{bmatrix}
    \boldsymbol{\varepsilon}_{xx} & \boldsymbol{\varepsilon}_{xy} & \boldsymbol{\varepsilon}_{xz} \\
    \boldsymbol{\varepsilon}_{yx} & \boldsymbol{\varepsilon}_{yy} & \boldsymbol{\varepsilon}_{yz} \\
    \boldsymbol{\varepsilon}_{zx} & \boldsymbol{\varepsilon}_{zy} & \boldsymbol{\varepsilon}_{zz} \\
    \end{bmatrix}
    =
    \begin{bmatrix}
    \boldsymbol{\varepsilon}_{yy} & -\boldsymbol{\varepsilon}_{yx} & -\boldsymbol{\varepsilon}_{yz} \\
    -\boldsymbol{\varepsilon}_{xy} & \boldsymbol{\varepsilon}_{xx} & \boldsymbol{\varepsilon}_{xz} \\
    -\boldsymbol{\varepsilon}_{zy} & \boldsymbol{\varepsilon}_{zx} & \boldsymbol{\varepsilon}_{zz} \\
    \end{bmatrix}.
\end{equation*}
It becomes evident that a crystal possessing a 90-degree rotational symmetry along the $z$-axis has specific zero elements in its dielectric tensor, as shown below:
\begin{equation*} 
\boldsymbol{\varepsilon} = 
    \begin{bmatrix}
    \boldsymbol{\varepsilon}_{xx} & \boldsymbol{\varepsilon}_{xy} & 0 \\
    -\boldsymbol{\varepsilon}_{xy} & \boldsymbol{\varepsilon}_{xx} & 0 \\
    0 & 0 & \boldsymbol{\varepsilon}_{zz} \\
    \end{bmatrix}.
\end{equation*}

\begin{figure}[t!]
    \centering
    \includegraphics[width=0.9\linewidth]{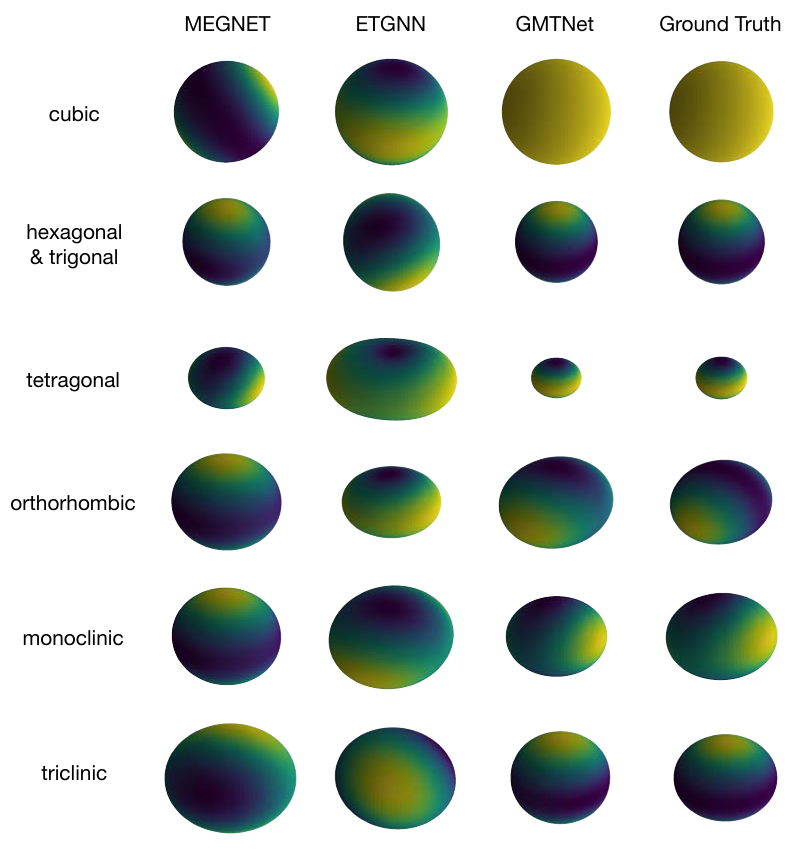}
    \caption{Comparative visualization of dielectric tensor predictions. This figure presents a model comparison for dielectric tensor prediction, on the dielectric test set comprising various crystal systems: cubic, hexagonal, trigonal, tetragonal, orthorhombic, monoclinic, and triclinic. GMTNet's predictions are highlighted for their alignment with the spatial symmetry characteristics of the ground truth tensors, underscoring its superior performance. In contrast, models such as MEGNET and ETGNN demonstrate a notable discrepancy in capturing these symmetry aspects. The comparison underscores GMTNet's effectiveness in generating symmetry-consistent dielectric tensor predictions.}
    \label{fig:vis_compare}
\end{figure}

\begin{figure}
    \centering
    \includegraphics[width=0.97\linewidth]{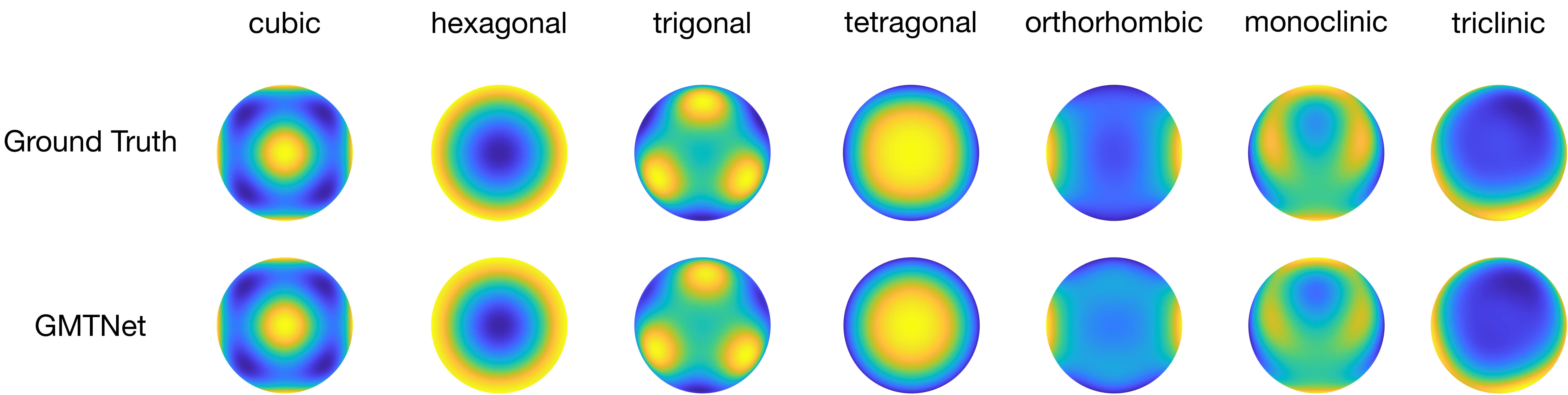}
    \caption{Visualization of elastic tensor predictions on elastic test set. The results underscore GMTNet's effectiveness in generating symmetry-consistent elastic tensor predictions  with tensor order 4.}
    \label{fig:vis_compare_elast}
\end{figure}

\subsection{GMTNet extracted crystal-level equivariant features meet the pre-established requirements}
\label{app:proof}

We provide analysis and proof in this section to demonstrate that GMTNet extracted crystal-level equivariant features meet the pre-established requirements in Sec.~\ref{sec:global_extract}.

\textbf{Assumptions}: crystal structure $\mathbf{M} = (\mathbf{A}$, $\mathbf{P}$, $\mathbf{L})$ satisfies given space group transformation $\mathbf{R} \in \mathbb{R}^{3\times3}, \mathbf{b} \in \mathbb{R}^{3}$ without any structural error, and the equivariant function $\theta_{equi}$ taking crystal structure $\mathbf{M} = (\mathbf{A}$, $\mathbf{P}$, $\mathbf{L})$ as input introduces no error including numerical ones.

\textbf{Proof}: for the transformation $\mathbf{R} \in \mathbb{R}^{3\times3}, \mathbf{b} \in \mathbb{R}^{3}$ in a given crystal's space group, since the crystal structure $\mathbf{M} = (\mathbf{A}$, $\mathbf{P}$, $\mathbf{L})$ satisfy this space group without structural error, we have the transformed crystal structure
$\mathbf{M}' = (\mathbf{A}$, $\mathbf{R}\mathbf{P} + \mathbf{b}$, $\mathbf{L})$, which is an equivalent crystal structure, albeit with a reindexed arrangement. Define the reindexed arrangement as a mapping function $\theta_{\text{re}}: i \to i', 1 \le i, i'\le n$. It can be seen that function $\theta_{\text{re}}$ is a bijection fucntion.

For the crystal-level equivariant feature after the transformation $\mathbf{R} \in \mathbb{R}^{3\times3}, \mathbf{b} \in \mathbb{R}^{3}$, by using the equivariant function $\theta_{equi}$, we can have
\begin{align*}
    \boldsymbol{f}_{i, \text{rotated}}^{\ell} = & \theta_{equi}(\mathbf{A}, \mathbf{R}\mathbf{P} + \mathbf{b}, \mathbf{L})^{\ell}_i \\
    = & \text{WD}^{\ell}(\mathbf{R}) \circ \theta_{equi}(\mathbf{A}, \mathbf{P}, \mathbf{L})^{\ell}_i \\
    = & \text{WD}^{\ell}(\mathbf{R}) \circ \boldsymbol{f}_{i}^{\ell},
\end{align*}
and
\begin{align*}
    \boldsymbol{G}^{\ell}_{\text{rotated}} =  \frac{1}{n} \sum_{1 \le i \le n} \boldsymbol{f}_{i, \text{rotated}}^{\ell} 
    = \frac{1}{n} \sum_{1 \le i \le n} \text{WD}^{\ell}(\mathbf{R}) \circ \boldsymbol{f}_{i}^{\ell} 
    = \text{WD}^{\ell}(\mathbf{R}) \circ \boldsymbol{G}^{\ell}.
\end{align*}

Further, by using the fact that $\mathbf{M}' = (\mathbf{A}$, $\mathbf{R}\mathbf{P} + \mathbf{b}$, $\mathbf{L})$ is an equivalent crystal structure as $\mathbf{M} = (\mathbf{A}$, $\mathbf{P}$, $\mathbf{L})$, albeit with a reindexed arrangement $\theta_{\text{re}}: i \to i', 1 \le i, i'\le n$, we can have
\begin{align*}
    \boldsymbol{G}^{\ell} = & \frac{1}{n} \sum_{1 \le i \le n} \boldsymbol{f}_{i}^{\ell} 
    = \frac{1}{n} \sum_{1 \le \theta_{\text{re}}(i) \le n} \boldsymbol{f}_{\theta_{\text{re}}(i)}^{\ell} = \frac{1}{n} \sum_{1 \le i' \le n} \boldsymbol{f}_{i'}^{\ell} = \boldsymbol{G}^{\ell}_{\text{rotated}},
\end{align*}
which means
\begin{align*}
    \boldsymbol{G}^{\ell} = \boldsymbol{G}^{\ell}_{\text{rotated}} = \text{WD}^{\ell}(\mathbf{R}) \circ \boldsymbol{G}^{\ell}.
\end{align*}
Hence, the proof that GMTNet extracted crystal-level equivariant features, detailed in Sec.~\ref{sec:global_extract}, meet the pre-established requirements and have the same spatial symmetry characteristics of the crystal structure is done.

\subsection{Directly converting outputs to satisfy crystal space group constraints results in sub-optimal performances}
\label{app:proof}

As discussed in the related work section, instead of carefully designing model components to achieve the invariance of crystal space group when predicting general tensors with different orders, there is a widely used general approach, group averaging, or more generally, frame averaging, that can convert any model output to be invariant to a group of transformations of interest. However, we show in this section that directly converting outputs to satisfy crystal space groups will result in sub-optimal prediction performances.

Specifically, we conduct experiments using group averaging with equation $\boldsymbol{\varepsilon} = \frac{1}{n_r} \sum \mathbf{R}_i \varepsilon \mathbf{R}_i^T$ to convert output of MEGNET and ETGNN to be space group invariant, and the results are shown in Tab.~\ref{tab:group_avg} where GA denotes using group averaging in the training and inference processes.

\begin{table}[h!]
    \vspace{-3mm}
  \caption{Comparison of accuracy in terms of Fnorm on the test set of dielectric tensors.}
  \small
  \label{tab:group_avg}
  \centering
  \begin{tabular}{l|cccc|c}
    \toprule
       & MEGNET & ETGNN & MEGNET-GA & ETGNN-GA & GMTNet \\
    \midrule
    Fnorm $\downarrow$& $4.16$& $3.92$& $4.51$& $4.07$  & $\mathbf{3.50}$ \\
    \bottomrule
  \end{tabular}
  \vspace{-2mm}
\end{table}

The results verify that directly converting outputs to satisfy crystal space groups will result in sub-optimal prediction performances, not only for methods that do not satisfy crystal tensor equivariance like MEGNET, but also for methods that satisfy crystal tensor equivariance like ETGNN.

\subsection{Experimental details and additional results}
\label{app:exp}

\textbf{Equivariance verification}. The equivariance of different models is verified on the dielectric test set using the $E(3)$ invariant measurement Fnorm shown in Table~\ref{tab_equi_verify}. ETGNN and our method maintain equivariance for tensor predictions while invariant MEGNET breaks this property.

\begin{table}[h!]
  \caption{Equivariance analysis in terms of Fnorm -- Evaluated on the most challenging triclinic system of dielectric tensors.}
  \small
  \label{tab_equi_verify}
  \centering
  \begin{tabular}{l|cc|c}
    \toprule
    Fnorm  & MEGNET & ETGNN & GMTNet \\
    \midrule
    Origin & 8.51 &8.32 & 7.65  \\
    Rotate 45-x & 9.74 &8.32 & 7.65   \\
    Rotate 45-y & 9.67 &8.32 & 7.65 \\
    Rotate 45-z & 8.51 &8.32 & 7.65  \\
    \midrule
    Equivariance & \ding{55} & \checkmark& \checkmark \\
    \bottomrule
  \end{tabular}
\end{table}

\textbf{Visualization comparison}. Fig.~\ref{fig:vis_compare} presents a comparative visualization of dielectric tensor predictions across various crystal systems within the dielectric test set. In our methodology, the visualization of the dielectric tensor is generated for each direction in three-dimensional space. This is achieved by considering a directional vector $\mathbf{v}_E = (x, y, z)$, normalized such that $||\mathbf{v}_E||_2 = 1$. We define the electric field vector as $\mathbf{E} = \mathbf{v}_E$ and subsequently compute the corresponding displacement response using the formula $\mathbf{D} = \boldsymbol{\varepsilon} \mathbf{E}$. To effectively represent the displacement response in every direction within the 3D space, we introduce a surface function $\theta_{\text{surface}}: (\mathbf{v}_E, \boldsymbol{\varepsilon}) \to \mathbf{D}$. The visualization is then rendered by utilizing the norm $||\mathbf{D}||_2$ to determine the radial distance $r$ from the origin to the surface and the surface color.

Notably, as shown in Fig.~\ref{fig:vis_compare}, GMTNet's predictions exhibit a high degree of alignment with the spatial symmetry characteristics inherent in the ground truth tensors, thereby highlighting its superior predictive capabilities. In contrast, baseline models such as MEGNET and ETGNN show large deviations in capturing these symmetry aspects. This comparative analysis shows GMTNet's effectiveness in producing dielectric tensor predictions that are consistent with the underlying symmetries of the crystal structures.

We further provide visualization comparison between ground truth piezoelectric and elastic tensors with GMTNet predicted ones on the piezoelectric and elastic test sets in Fig.~\ref{fig:vis_compare_piezo} and Fig.~\ref{fig:vis_compare_elast} using MTEX~\cite{mtex}. These comparisons further demonstrate GMTNet's effectiveness in generating symmetry-consistent higher order tensor predictions with tensor order 3 (piezoelectric tensors) and 4 (elastic tensors).

\begin{figure}[t!]
    \centering
    \includegraphics[width=0.95\linewidth]{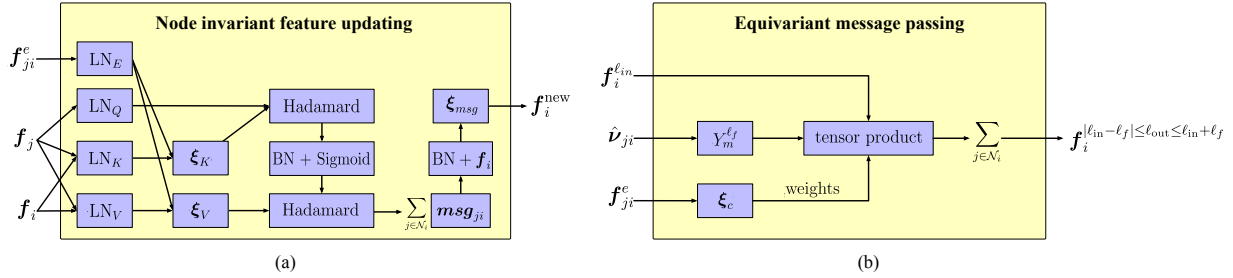}
    \caption{Message passing details for node invariant feature updating and equivariant message passing in GMTNet. (a) Node invariant feature updating that takes node features ($\boldsymbol{f}_j$, $\boldsymbol{f}_i$) and edge feature ($\boldsymbol{f}_{ji}^e$) as input to obtain the updated invariant node feature $\boldsymbol{f}_i^{\text{new}}$. (b) Equivariant message passing that produces high rotation order node features.
    In this figure, all notations follow Sec.~\ref{sec:global_extract}.}
    \label{fig:arch}
\end{figure}

\textbf{Configurations of GMTNet}. The detailed architecture of node invariant feature updating and equivariant message passing within GMTNet is shown in Fig.~\ref{fig:arch}. For the edge embedding, we use $c=0.75$ and RBF kernels with value from $-4$ to $0$ which maps $-c/||\boldsymbol{\nu}_{ji}||_2$ to a 512 dimensional vector. This 512 dimensional vector is mapped to 128 dimensional vector by a nonlinear layer to $\boldsymbol{f}_{ji}^e$. We use 2 layers of node invariant feature updating and 3 layers of equivariant message passing for all tasks including dielectric, piezoelectric, and elastic tensors. Learning rate of 0.001, epoch number of 200, and batch size of 64 are used for dielectric, piezoelectric, and elastic tensors. We construct crystal graphs using radius determined by the 16-$th$ nearest neighbor. The source code of GMTNet will be released when the paper is publicly available. 

\textbf{Implementations of MEGNET}. Following the original paper~\cite{megnet}, we use three layers of MEGNET message passing with the same feature dimensions as mentioned in the paper. We train MEGNET for 200 epochs using Huber loss with a learning rate of 0.001 and AdamW optimizer with $10^{-5}$ weight decay. The same polynomial learning rate decay scheduler as our GMTNet implementation is used. To predict a tensor matrix of shape $3\times 3$, we change the original output dimension from one to nine.

\textbf{Implementations of ETGNN}. Following ETGNN~\cite{etgnn}, we use four DimeNet++ layers with hidden dimension 128 and ELU activation function to serve as the invariant message-passing network. Since the code of ETGNN is not publically accessible, we implement DimeNet++ layers following the code provided by \citet{du2023m}. We train ETGNN for 200 epochs using Huber loss with a learning rate of 0.001 and AdamW optimizer with $10^{-5}$ weight decay. The same polynomial learning rate decay scheduler as our GMTNet implementation is used.

\textbf{Tolerance-guided prediction adjustment}. As discussed in Sec.~\ref{sec:settings}, the Wigner D matrix computation provided in the e3nn package~\cite{e3nngeiger2022}, introduces numerical errors, notably more pronounced in features with higher rotation orders. To mitigate the impact of this issue, we implement a tolerance-guided prediction adjustment during inference.

The symmetry adjustment operator, $\frac{1}{{n_{r}}}\sum_{1\le i \le {n_{r}}}\text{WD}^{\ell} (\mathbf{R}_i)$, embeds the structural symmetry information of the crystal input. However, the minor errors in each element are challenging to identify and correct. Fortunately, for specific tensor properties, it is possible to extract symmetry characteristics using defined tolerances, thereby eliminating errors in the symmetry operator.

For instance, consider the symmetry adjustment operator $\frac{1}{{n_{r}}}\sum_{1\le i \le {n_{r}}}\text{WD}^{\ell=0,1,2} (\mathbf{R}_i)$. To extract the equality characteristics of dielectric tensors using tolerance, we first construct three vectors $\mathbf{v}_1 \in \mathbb{R}, \mathbf{v}_2 \in \mathbb{R}^3, \mathbf{v}_3 \in \mathbb{R}^5$ with rotation orders 0, 1, and 2, respectively, each containing distinct values in their elements. The symmetry-adjusted vectors are obtained as $\mathbf{v}_1^{\text{new}} = \frac{1}{{n_{r}}}\sum_{1\le i \le {n_{r}}}\text{WD}^{\ell=0} (\mathbf{R}_i) \circ \mathbf{v}_1, \mathbf{v}_2^{\text{new}} = \frac{1}{{n_{r}}}\sum_{1\le i \le {n_{r}}}\text{WD}^{\ell=1} (\mathbf{R}_i) \circ \mathbf{v}_2, \mathbf{v}_3^{\text{new}} = \frac{1}{{n_{r}}}\sum_{1\le i \le {n_{r}}}\text{WD}^{\ell=2} (\mathbf{R}_i) \circ \mathbf{v}_3$. Since dielectric tensors comprise three vectors of rotation orders 0, 1, and 2, the vectors $\mathbf{v}_1^{\text{new}}, \mathbf{v}_2^{\text{new}}, \mathbf{v}_3^{\text{new}}$ can be transformed into a $3 \times 3$ dielectric tensor $\boldsymbol{\varepsilon}_{\text{symmetry}}$. Equality pairs within $\boldsymbol{\varepsilon}_{\text{symmetry}}$ are identified by comparing elements with a tolerance threshold. We find that using 0.01\% of the mean value of two elements as the threshold is effective for determining equality in dielectric tensors. Similar approaches can be applied to piezoelectric and elastic tensors, and threshold value $\epsilon_{\text{zero}}=1.0$ is effective for determining zero elements in piezoelectric and elastic tensors if needed.

\end{document}